\documentclass{llncs}
\usepackage{amsmath,amssymb,amscd}

\DeclareMathOperator\nfm{nfm}
\DeclareMathOperator\pfm{pfm}
\DeclareMathOperator\compliance{\ast\mspace{-11mu}\sim}

\begin{document}
\bibliographystyle{splncs}

\title{The Separation of Duty with Privilege Calculus 
	\thanks{We are grateful for the supporting of the National Natural 
		Science Foundation of China (NSFC, Project No. 70401001).}
}

\titlerunning{Privilege Calculus}

\author{Chenggong Lv\inst{1} 
	\and Jun Wang\inst{1} 
	\and Lu Liu\inst{1} 
	\and Weijia You\inst{1}}
\institute{Beihang University, Beijing 100083, P.R. China, \\ 
	\email{lcgong@gmail.com},
    \email{king.wang@buaa.edu.cn}, \\
	\email{liulu@buaa.edu.cn},
	\email{weijiawx@gmail.com}}

\maketitle

\begin{abstract}
This paper presents Privilege Calculus (PC) as a new approach of
    knowledge representation for Separation of Duty (SD)
    in the view of process
    and intents to improve the reconfigurability and traceability of SD.
PC presumes that 
    the structure of SD should be reduced to the structure of privilege 
    and then the regulation of system should be analyzed 
        with the help of forms of privilege.
\end{abstract}

\section{Introduction}

The Separation of Duty (SD) is a security principle
    that is used to formulate multi-person control policies,
    which requires that two or more different people be responsible for
    completion of a task or a set of related tasks \cite{simon1997sdr}.
The Role-Based Access Control (RBAC) system is defined by a state machine model
    and characterized by the fact that
    a user's rights to access objects are defined
        by the user's membership to a ``role''
        and by the roles' permissions to perform operations on those objects
        \cite{gligor1998fds}.
Hence, the role is a semantic referent of duty representation 
    and the structure of role is a division of
    rights in cross-organization systems.
With the help of assignment operation,
    the user-role assignment can be handled by one
    while permission-role assignment is handled by another \cite{sandhu2001fdr}.

Because the permission assignment on role hierarchy is static,
    Sandhu \cite{sandhu1998rah} introduced the Role Activation Hierarchy (RAH).
RAH extends the permission-usage hierarchy
    and makes the role activation governed by an activation hierarchy.
Sandhu argued that the administration of RBAC must itself be decentralized
    and managed by administrative roles.
Moreover, Ferraiolo \cite{ferraiolo2001pns} argued that
    static separation of duty enforces constraints on the assignment of users to roles,
    and dynamic separation duty places constraints on roles that
        can be activated within or across a user's session.

Although the delegation model \cite{barka2000frb} is helpful to
    resolve the temporal permission assignment problem
        by the delivery of duty in trust,
    the permission delegated has to crosscut two or more roles in RAH
    and the definition working to map between them is not easy.
Also, for the constraints in RBAC,
    there is an inconsistency between the access control policy
		and the constraints that are specified to limit this policy.
One  transform  limit may  preclude,  by  a  constraint,
    the  change  in  another  transform  limit  even  though  the  rights
        that embody  the  conflict  have  not  been  assigned  yet \cite{jaeger1999iic}.
So extra mechanisms were integrated to detect \cite{schaad2001dcr}
	and resolve \cite{jaeger2004rcc} the conflict.
Jaeger has argued that
	since  fail-safety  is  often  a  goal  of  secure  systems,
	some  form  of  conflict  resolution  may  not  be  unreasonable,
	but  the  trade-off  is  not  clear-cut \cite{jaeger1999iic}.

It is the question that
	how to keep change of condition predictable 
	and how control exists after reconfiguration in dynamic way,
    for which the essential challenge is, 
        we believe , the representation of SD still.
Our approach is enlightened by $\pi$-calculus 
	that makes process reconfigurable \cite{milner1999cam},
    and assumes that
    the duty is composed of the interaction commitment of process, 
	i.e. privilege(see section \ref{sec:privilege}),
	and the result of SD is a collection of interaction commitments,
	i.e. regulation(see section \ref{sec:regulation}).
The examples in section \ref{sec:disccussion} show 
    the flexibility and usefulness of our approach.

\section{Regulation} \label{sec:regulation}

There are two synchronized complementary actions in an interaction 
	\cite{milner1999cam}.
The guarded action is an action with one preceding action
    that has not been reduced.
We have two processors that execute these actions respectively. 
These actions represent 
	the semantics of this interaction of the two processors.

A component is featured with the composition of distinct functions 
	and consists of corresponding processors.
One function features one processor in design,
    and one processor runs one action in one process (runtime).
The sequence of {\em observed} action  represents a process 
	and reflects the implementation of function intention.
So the sequence of {\em programmed} action
	represents an interaction commitment.
Moreover, the intersection of interaction commitment 
	involved in an interaction are not empty.

Although component is neutral, system works in a conservative way.
The framework of system is a guarding processor
	and guards each interaction of two managed components.
The guarding interaction of framework 
	precedes the guarded interaction of component.

Regulation of system is a collection of interaction commitments,
	including the interaction commitments of framework and of component.
For the systems based on privilege calculus, 
	the result of separation of duty is regulation, 
	i.e. a collection of privilege.

\section{Structure of Privilege}
In this section, we give the structure of privilege 
	with the help of notions, employment and condition.
The notion of employment is the refined structure of function intention.

\subsection{Employment}

\begin{definition}
The function-entity employment $f/e$ means that 
    function $f$ is employed on entity $e$.
\end{definition}

\begin{proposition}
There are employments, $f_1/e_1$ and $f_2 / e_2$, \[
f_1 / e_1 + f_2/e_2 = \emptyset
    \iff f_1 / e_1 = \emptyset \wedge f_2 / e_2 = \emptyset
\]
\end{proposition}
Then we introduce the left employment mergence of function-entity.
\begin{proposition}
There are employments, $f_1/e_1 \not= \emptyset$ and $f_2 / e_2 \not= \emptyset$. \[
(f_1 / e_1)*(f_2 / e_2) =
\begin{cases}
  f/e, &\text{if $f=f_1=f_2 \not=\emptyset$ and $e=e_1=e_2 \not=\emptyset$;} \\
  \emptyset , &\text{otherwise.}
\end{cases}
\]
\end{proposition}

\begin{definition}
\label{employment}
$F$ is a collection of functions, and $E$ is a collection of entities.
The employment $F/E$ is a set $\left\{ f/e | f \in F, e \in E \right\} $.
\end{definition}

Let $F$, $F_1$, and  $F_2$ be respectively a collection of functions,
    and let $E$, $E_1$, and $E_2$ be a collection of entities.
    We have $f_1 \in F_1$, $f_2 \in F_2$, $e_1 \in E_1$, and $e_2 \in E_2$.
The mergence of employment is
\begin{equation}
\label{eq:merge}
F_1/E_1 * F_2/E_2
    = \left\{ f_1/e_1 * f_2/e_2 \not= \emptyset \right\} \ .
\end{equation}
The composition of employment is
\begin{equation}
\label{eq:compos}
    F_1/E_1 + F_2/E_2 =
        \left\{ f_1/e_1 \not= \emptyset
            \vee f_2/e_2 \not= \emptyset \right\} \ .
\end{equation}

For the convenience of computation, we give
    $F / \emptyset = \emptyset$, $\emptyset / E = \emptyset$
    and $\emptyset / \emptyset = \emptyset$.
If no confusion arises, these expressions,$f/e$,  $\left\{ f \right\} / e$ and $f / \left\{e\right\}$, are the same as $\left\{f\right\} / \left\{e\right\}$.
With definition \ref{employment}
    and equations \ref{eq:merge} and \ref{eq:compos},
    we prove that the employment are associative, commutative and distributive.

\subsection{Condition}
Regulation is different from process,
    which we have discussed in section \ref{sec:regulation}.
The condition acts as the connection with the state of ``process world''.
In this subsection, we propose the definition of condition.

\begin{definition}
The fact set $T$ is a collection of subsets of statement collection $S$.
The fact set $T$ on $S$ has the following properties:
\begin{enumerate}
    \item $\emptyset$ and $S$ are in $T$.
    \item The union of the elements of any sub-collection of $T$
        is in $T$.
    \item The intersection of the elements of any finite sub-collection of $T$ in $T$.
\end{enumerate}
\end{definition}

\begin{definition}
Fact set $T$ on $S$, condition $r$ is a function
    $r:T_s \rightarrow \left\{ 1,0 \right\}$
    with the property: 
	$ \forall x_1$, $x_2 \in T$ and $ x_1 \cap x_2 = \emptyset$,
    	$ r(x_1 \cup x_2) = r(x_1) \vee r(x_2) $.
\end{definition}
The $\left\{ 1,0 \right\}$ is the true value. If the fact $x \in T$,
    we call that the condition $r$ is supported on the fact $x$,
    or the fact $x$ supports the condition $r$.

\begin{proposition}
For fact set $T$ on $S$, $\forall x_1, x_2 \in T$ and $x_1 \subset x_2$,
    $r(x_1) \rightarrow r(x_2)$ .
\end{proposition}

\begin{definition}
For fact set $T$ on $S$ and condition $r$, if $r(x)$ is true,
    the fact $x \in T$ is the evidence to $r$.
\end{definition}

\begin{definition}
For fact set $T$ on $S$,
    $\exists x^* \in T$ and such that
        $x^*$ is the evidence to the condition $r$,
    if $\nexists x \subset x^*$ and such that $x$ is the evidence to $r$,
    then the $x^*$ is the minimum evidence to $r$.
\end{definition}

%-----------------------------------------------------------------------------
\subsection{Privilege} \label{sec:privilege}

\begin{definition}
For a collection of functions $F$, a collection of entities $E$
    and a collection of conditions $R$,
    the privilege is $(F/E,R)$.
\end{definition}
For convenience,
    we define, $(\emptyset, r) = \emptyset$.

\begin{definition}
\label{def:priv}
The privilege space $\mathcal{P}$ is a collection of subsets of $P$
    with the following properties:
\begin{enumerate}
   
\item {\em (Privilege Mergence)} For all privilege, 
    $u,v\in\mathcal{P}$, $u=(f_1/E_1,R_1)$, and $v=(f_2/E_2,R_2)$,
    $$ u*v =\{(f_1 * f_2 / (E_1 \cap E_2), R_1 \cap R_2) \} \ ;$$

\item {\em (Privilege Composition)} For all privilege,
    $u,v\in\mathcal{P}$, $u=(f_1/E_1,R_1)$, and $v=(f_2/E_2,R_2)$,
    $$u+v =\{(f_1/E_1,R_1) \cup (f_2/E_2,R_2) \} \ ;$$

\item For all privilege, $u,v\in\mathcal{P}$, 
    $u*v=v*u$;

\item For all privilege, $u,v\in\mathcal{P}$, 
    $u+v=v+u$;

\item For all privilege, $u,v,w\in\mathcal{P}$, 
    $(u*v)*w = v*(u*w)$;

\item For all privilege, $u,v,w\in\mathcal{P}$, 
    $(u+v)+w = v+(u+w)$;

\item For all privilege, $u,v,w\in\mathcal{P}$, 
    $u*(v+w) = u*v+u*w$.

\end{enumerate}

\end{definition}

%---------------------------------------------------------------------------
\section{Normal Form of Privilege}
\begin{definition}
The employment arrangement $M$ is a finite collection of employment 
    and such that
    $ \forall m, n \in M, \quad  m\not=n \wedge m*n=\emptyset$.
\end{definition}
\begin{definition}
To employment arrangement $M$, the normal form of privilege $p$ is
$$ \nfm_{M}(p) = \sum_{i}^{M} {m_i} = \sum_{i}^{M} (f_{i}/E_{i},c_{i}) \ ,$$
where $f_i/E_i$ is an element of $M$ and $c_i$ is a condition.
\end{definition}

\begin{proposition}
To employment arrangement $M$, every privilege is structurally equal
    to its normal form.
\end{proposition}

\begin{definition}
To employment arrangement $M$, the privileges are structural equivalence,
    if and only if they have the same normal form,
    $$ u \stackrel{M}{=} v \iff \nfm_M(u) = \nfm_M(v) \ .$$
\end{definition}

When one condition has an evidence,
    these privileges that involve the condition are pulsed.
Corresponding to normal form of privilege,
    there is the pulsed form.
\begin{definition}
\label{def:pulse}
To employment arrangement $M$, on the fact $t \in T$,
    the pulsed form of privilege $p$ is
    $$ \pfm_M(p,t) = \sum_{i}^{M} (f_{i}/E_{i}, c_i(t)) \ , $$
where $f_i/E_i$ is an element of $M$ and $c_i$ is a condition.
\end{definition}

We have a sequence of fact $ Q = (t_0, t_1, \dots, t_j, \dots) $.
    We get the sequence of pulse to privilege $t$, \[
    \pfm_M(p,Q) = (\pfm_M(p,t_0), \pfm_M(p,t_1), \dots , \pfm_M(p,t_j),\dots)
    \ .\]
This sequence of pulsed form describes
    the trace of process about privilege $p$.
The trace matrix $(c_{i,j})$ of privilege $p$ is made from this sequence,
    where $c_{i,j} \in \left\{ 1,0 \right\}$ .
\[
    \begin{array}{c|cccccc}
            & t_0    & t_1      & \dots   & t_j     & \dots   \\
    \hline
    f_0/E_0 &c_{0,0} & c_{0,1}  & \dots   & c_{0,j} & \dots   \\
    f_1/E_1 &c_{1,0} & c_{1,1}  & \dots   & c_{1,j} & \dots   \\
    \vdots  &\vdots  & \vdots   & \ddots  & \vdots  & \ddots  \\
    f_i/E_i &c_{i,0} & c_{i,1}  & \dots   & c_{i,j} & \dots   \\
    \vdots  &\vdots  & \vdots   & \ddots  & \vdots  & \ddots  \\
    f_n/E_n &c_{n,0} & c_{n,1}  & \dots   & c_{n,j} & \dots
    \end{array}
\]

For example, we have two operations (privileges) $op_1$ and $op_2$,
    and three people (privileges) $u_1$, $u_2$ and $u_3$.
We want to know what will happen at time (facts) $t_0$ and $t_1$.
So we define a gauging privilege, 
    $ g = (u_1 + u_2 + u_3) * (op_1 + op_2)$.
And the sequence of pulse is $(\pfm_M(g,t_0), \pfm_M(g,t_1)) $.

\begin{definition}
To employment arrangement $M$,
    privileges, $u$ and $v$, are congruent on fact $t \in T$,
    $a \stackrel{t}{\sim} b $,
    if and only if $u$ and $v$ have the same pulsed form.
\end{definition}

\begin{definition}
To employment arrangement $M$, on fact $ t \in T $,
    privilege $p$ is compliant to privilege $q$,
    $ p \stackrel{t}{\compliance} q $,
    if and only if $ (p*q) \stackrel{t}{\sim} q $.
\end{definition}

The congruence $\sim$ and the compliance $\compliance$ are a function
    $ P \times P \times T \rightarrow \left\{ 1,0 \right\} $.
So they can be a condition in one high-order privilege.
For a compliance example, we have the privileges, $g$, $p$ and $q$, 
	and such that $g = \left[p \compliance q \right]$.
We call that the privilege $g$ is a high-order privilege of $p$ and $q$.

\section{Discussion}  \label{sec:disccussion}

In general, the role-based models,
    such as RBAC reference model \cite{sandhu1996rba,ferraiolo2001pns},
        ARBAC \cite{sandhu1999amr}, and T-RBAC \cite{oh2003trb},
    have constructs,
        such as, USERS, ROLES, OPS (operations), and OBJS (objects),
    and relations,
        such as UA(user-to-role assignment), PA(permission-to-role assignment), PRMS (set of permission), and RH (role inheritance relation).
These constructs are able to be defined with privilege and these relation with privileges.
And these privileges are glued by privilege's operations,
    such as privilege mergence and privilege composition.

The following code is a demonstration written in PAL(Privilege Analysis Language)
    that is a reference implementation based on privilege calculus.
With this demonstration we discuss cases about privilege representation.

\begin{verbatim}
    namespace "example" {

      let doc1 is TechDoc

      reader := (read + list)/TechDoc
      manager := (reader + write + remove)/TechDoc

      bob := reader + write/TechDoc
      may := manager

      phone := read + list
      officepc := read + list + write + remove
    }
\end{verbatim}

Shown by the above code, we have
    four operations, $read$, $list$, $write$, and $remove$,
    two roles, $reader$ and $manager$,
    two users, $bob$ and $may$,
    and two terminals, $officepc$ and $phone$.
The statement ``$let$'' declares that $doc1$ is a document in the category $TechDoc$.
The role $reader$ can read any documents in $TechDoc$ and $list$ entries of those,
    and the role $manager$ can $write$ and $remove$ any one in $TechDoc$
    and $manager$ inherits all of $reader$'s privileges that are limited in $TechDoc$.
User $bob$ plays the role $reader$
     and User $may$ has the role $manager$.
The mobile $phone$, a terminal device,
    has a limitation to access, $read$ and $list$.

So far, we have defined these privileges:
    $read$, $list$, $write$, $remove$,
    $reader$, $manager$, $bob$, $may$,
    $of{f}icepc$, $phone$, $doc1$, and $TechDoc$.

While user $bob$ has logged in system at his $of{f}icepc$,
    and the system creates his session, $session_1 = bob * officepc $.
In $session_1$,
    $bob$ is able to $read$, $list$ and $write$ any one in $TechDoc$.

Later $bob$ uses his personal $phone$ to navigate the system,
    the $session_2$ is created automatically, $session_2 = bob * phone$ .
The $session_2$'s privileges are different from $session_1$'s.
We set an employment arrangement, $ M = read + list + write + remove$.
Thus,
\begin{align*}
session_1 &\stackrel{M}{=}  bob * of{f}icepc \\
    & \stackrel{M}{=}  (reader + write/TechDoc) * (read + list + write + remove) \\
    & \stackrel{M}{=} read/TechDoc + list/TechDoc + write/TechDoc \ ,\\
session_2  &\stackrel{M}{=} bob * phone \\ 
    &\stackrel{M}{=} ((read + list)/TechDoc + write) * (read + list) \\
    &\stackrel{M}{=} read/TechDoc + list/TechDoc \ .
\end{align*}
With the above computation,
    we know the $session_2$ lacks the employment `$write$' on $TechDoc$.
It is interesting that the session in system can be created as a privilege
    and these constructs,
        such as session, user, role, permission, group, location etc.,
    could be represented by privilege.

We continue the story.
User $bob$ wants to read the document $doc1$ that is a $TechDoc$.
The guard $readguard$ to the action $read$ is
\[
    readguard = read * \left[ session_1 \compliance \left( read/doc1 \right) \right]
\ .\]
The $readguard$ is the high-order privilege of $session_1$ and $read/doc1$.
The pulse of $readguard$ depends on the $session_1$'s compliance to $read/doc1$.

User $may$ has logged in, and her session is $session_3$.
She wants to write the document $doc1$.
The regulation does concern not only $may$'s privilege but also the $doc1$'s.
So the privilege $doc1$ is redefined, $  doc1 = readable + writable$.
Because the $doc1$'s ``writable'' action
    and the $may$'s ``write'' action are complementary in this synchronized interaction,
    $writeguard$ and $writableguard$ are defined,
\begin{align*}
    writegurad &= write * \left[session_3 \compliance (write/doc1) \right] \ ,\\
    writableguard &= writable * \left[doc1 \compliance (writable) \right] \ .
\end{align*}
Thus, we have the interaction guard $interactionguard$,
\[  interactionguard = writeguard + writableguard  \ .\]
Finally, the $session_3$'s compliance and the $doc1$'s compliance
    consistently make the pulse of $interactionguard$.

\section{Conclusion}

Separation of duty is critical not only in security control 
	but also in modeling and monitoring of business logic.
For improving reconfigurability of representation of duty,
	we propose {\em privilege calculus}.
With the help of privilege's normal form and pulsed form,
    we are able to analyze the structure of privilege
    and to monitor the change in process.
We also have demonstrated that
    the access control model based on privilege calculus
        is compatible with RBAC, ACL.

So far, we have only begun to explore
    the computation of privilege and representation of regulation
    in access control logic.
But we have little knowledge about the relationship
    among regulation, business process and business rule. 
On all accounts, we hope that
    the paper will throw some light on
        the knowledge representation in separation of duty domain
    to facilitate the analysis of business rules and business processes.

\end{document}